\documentclass[a4paper]{article}

\usepackage[english]{babel}
\usepackage[utf8x]{inputenc}
\usepackage[T1]{fontenc}

\usepackage[a4paper,top=3cm,bottom=2cm,left=3cm,right=3cm,marginparwidth=1.75cm]{geometry}

\usepackage{amsmath}
\usepackage{graphicx}
\usepackage[colorinlistoftodos]{todonotes}
\usepackage[colorlinks=true, allcolors=blue]{hyperref}

\usepackage{amsfonts}
\usepackage{mathtools}
\usepackage{amssymb}

\newcommand{\transpose}{^{\operatorname{T}}}

\newcommand{\xinitial}{x_I}

\newcommand{\xtarget}{x_T}

\newcommand{\bmc}{}
\newcommand{\emc}{}
\renewcommand{\bmc}{\begin{multicols}{3}}
\renewcommand{\emc}{\end{multicols}}

\usepackage{algorithm,algorithmicx}
\DeclareMathOperator*{\minimize}{\text{minimize}}

\DeclareMathOperator*{\stj}{\text{subject to}}

\newcommand{\transposeSR}{\intercal}
\usepackage{authblk}
\newcommand{\beginsupplement}{%
        \setcounter{table}{0}
        \renewcommand{\thetable}{S\arabic{table}}%
        \setcounter{figure}{0}
        \renewcommand{\thefigure}{S\arabic{figure}}%
     }

\title{An Algorithm for Cellular Reprogramming}
\author[1]{Scott Ronquist}
\author[2]{Geoff Patterson}
\author[3]{Markus Brown}
\author[1]{Stephen Lindsly}
\author[1]{Haiming Chen}
\author[4]{Lindsey A. Muir}
\author[5]{Max Wicha}
\author[6]{Anthony Bloch}
\author[7]{Roger Brockett}
\author[1,4,8]{Indika Rajapakse}
\affil[1]{Department of Computational Medicine and Bioinformatics, Medical School, University of Michigan, Ann Arbor, MI 48109}
\affil[2]{IXL Learning, Raleigh, NC 27560}
\affil[3]{Department of Biological Sciences, University of Maryland, College Park}
\affil[4]{Department of Pediatrics and Communicable Diseases, University of Michigan, Ann Arbor, MI 48109}
\affil[5]{Department of Hematology/Oncology, University of Michigan, Ann Arbor, MI 48109}
\affil[6]{Department of Mathematics, University of Michigan, Ann Arbor, MI 48109}
\affil[7]{Harvard School of Engineering and Applied Sciences, Cambridge, MA 02138}
\affil[8]{Corresponding author: indikar@umich.edu}
\date{}
\begin{document}
\maketitle

\begin{abstract}
The day we understand the time evolution of subcellular elements at a level of detail comparable to physical systems governed by Newton's laws of motion seems far away. Even so, quantitative approaches to cellular dynamics add to our understanding of cell biology, providing data-guided frameworks that allow us to develop better predictions about, and methods for, control over specific biological processes and system-wide cell behavior. In this paper, we describe an approach to optimizing the use of transcription factors (TFs) in the context of cellular reprogramming. We construct an approximate model for the natural evolution of a cell cycle synchronized population of human fibroblasts, based on data obtained by sampling the expression of 22,083 genes at several time points along the cell cycle. In order to arrive at a model of moderate complexity, we cluster gene expression based on the division of the genome into topologically associating domains (TADs) and then model the dynamics of the TAD expression levels. Based on this dynamical model and known bioinformatics, such as transcription factor binding sites (TFBS) and functions, we develop a methodology for identifying the top transcription factor candidates for a specific cellular reprogramming task. The approach used is based on a device commonly used in optimal control. Our data-guided methodology identifies a number of transcription factors previously validated for reprogramming and/or natural differentiation. Our findings highlight the immense potential of dynamical models, mathematics, and data-guided methodologies for improving strategies for control over biological processes.
\end{abstract}





\section*{Significance Statement}
Reprogramming the human genome toward any desirable state is within reach; application of select transcription factors drives cell types toward different lineages in many settings. We introduce the concept of data-guided control in building a universal algorithm for directly reprogramming any human cell type into any other type. Our algorithm is based on time series genome transcription and architecture data and known regulatory activities of transcription factors, with natural dimension reduction using genome architectural features. Our algorithm predicts known reprogramming factors, top candidates for new settings, and ideal timing for application of transcription factors. This framework can be used to develop strategies for tissue regeneration, cancer cell reprogramming, and control of dynamical systems beyond cell biology.

\section*{Introduction}
In 1989, pioneering work by Weintraub \emph{et al.} successfully reprogrammed human fibroblast cells to muscle cells via over-expression of transcription factor (TF) MYOD1, becoming the first to demonstrate that the natural course of cell development could be altered \cite{Weintraub1989}. In 2007, Yamanaka \emph{et al.} changed the paradigm further by successfully reprogramming human fibroblast cells to an embryonic-stem-cell-like state (induced pluripotent stem cells; iPSCs) using four TFs: POU5F1, SOX2, KLF4, MYC. This showed that a differentiated cell state could be reverted to a more pluripotent state \cite{Yamanaka-2007}.

These remarkable findings demonstrated that the genome is a system capable of being controlled via an external input of TFs. In this context, determining how to push the cell from one state to another is, at least conceptually, a classical problem of control theory \cite{Brockett}. The difficulty arises in the fact that the dynamics -- and even proper representations of the cell state and inputs -- are not well-defined in the context of cellular reprogramming. Nevertheless, it seems natural to treat reprogramming as a problem in control theory, with the final state being the desired reprogrammed cell. In this paper, we provide a control theoretic framework based on empirical data and demonstrate the potential of this framework to provide novel insights into cellular reprogramming \cite{Rajapakse-2011,doublehelix}.
 
Our goal is to mathematically identify TFs that can directly reprogram human fibroblasts to a desired target cell type. As part of our methodology, we create a model for the natural dynamics of proliferating human fibroblasts. We couple data from bioinformatics with methods of mathematical control theory--a framework which we dub \emph{data-guided control} (DGC). Using time series data and a natural dimension reduction through topologically associating domains (TADs), we capture the natural dynamics of the cell, including the cell cycle.

We use this model to determine a principled way to identify the best TFs for efficient reprogramming of a given cell type toward a desired target cell type. Previously, selection of TFs for reprogramming has been based largely on trial and error, typically relying on TF differential expression between cell types for initial predictions. Recently, Rackham \emph{et al.} devised a predictive method based on differential expression as well as gene and protein network data \cite{rackham-2016}. Our approach is fundamentally different in that we take a dynamical systems point of view, opening avenues for investigating efficiency (probability of conversion), timing (when to introduce TFs), and optimality (minimizing the number of TFs and amount of input).

Using genomic transcription and architecture data, our method identifies TFs previously found to reprogram human fibroblasts into embryonic stem cell-like cells and reprogram fibroblasts into muscle cells. Our method also predicts TFs for conversion between human fibroblasts and many additional target cell types. 
In addition, we show the efficacy of using TADs for genome dimension reduction. Our analysis predicts the points in the cell cycle at which the insertion of TFs can most efficiently affect a desired change of cell state. Implicit in this approach is the notion of distance between cell types, which is measured in terms of the amount of transcriptional 
\textcolor{black}{change} 
required to transform one cell type into another. In this way, we are able to provide a comprehensive quantitative view of human cell types based on the respective distances between cell types.

Our framework separates into three parts:

\begin{enumerate}
\item \textbf{Define the state.} Use structure and function observations of the initial and target cell types' genomes to define a comprehensive state representation.
\item \textbf{Model the dynamics.} Apply model identification methods to approximate the natural dynamics of the genome, from time series data.
\item \textbf{Define and evaluate the inputs.} Infer from bioinformatics (TF binding location and function) where TFs can influence the genome, then quantify controllability properties with respect to the target cell type.
\end{enumerate}

The actual dynamics of the genome are undoubtedly very complicated, but as is often done in mathematical modeling studies, we use measurements to identify a linear approximation. This will take the form of a difference equation that is widely studied in the control systems literature, \cite{astrom2010feedback}
\begin{equation}
x_{k+1}=A_k x_k+Bu_k. \label{eqn-linear}
\end{equation}
\noindent In this case, the three items listed above correspond respectively to the value of the  state $x_k$ at time $k$, the time dependent state transition matrix $A_k$, and the input matrix $B$ (along with the input function $u_k$).

\section*{Methods}
\subsection*{Genome State Representation and Dimension Reduction: $\mathbf x_k$}

The state representation $x$ in Eq. \ref{eqn-linear} is the foundation for any control system and is critical for controllability analysis. To fully represent the state of a cell, a high number of measurements would need to be taken, including gene expression, protein level, chromatin conformation, and epigenetic measurements. As an initial simplification, we assume that the gene expression profile is a sufficient representation of the cell state. 

\begin{figure*}[h]
\centering
\includegraphics[width = .5\textwidth]{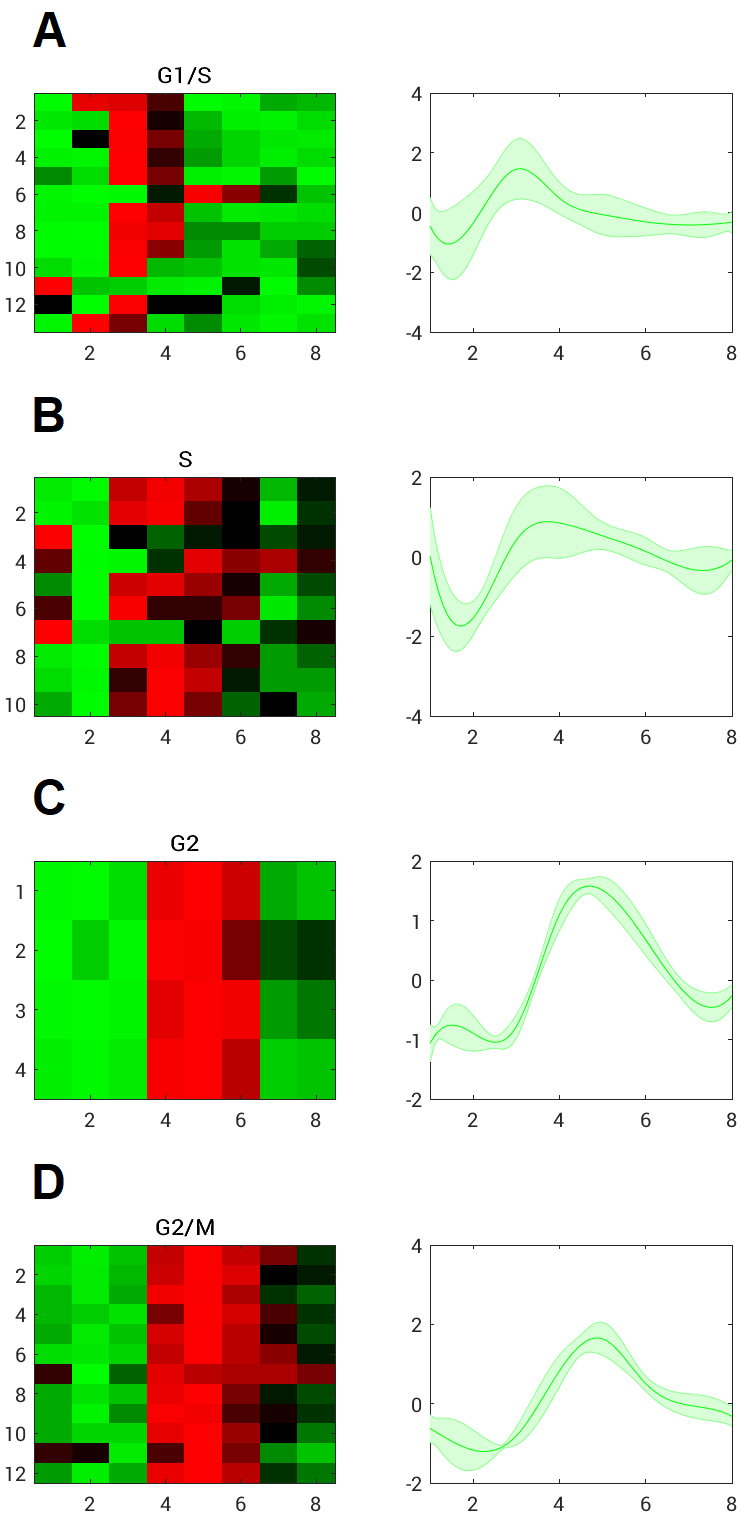}
\caption{Analysis of cell cycle marker gene expression. Gene expression RNA-seq data for 39 genes that have been shown in the literature to be cell cycle regulated \cite{Whitfield-2002}. Cell cycle phases shown include (\textit{A}) \texorpdfstring{G\textsubscript{1}}{G1}/S, (\textit{B}) S, (\textit{C}) \texorpdfstring{G\textsubscript{2}}{G2}, (\textit{D}) \texorpdfstring{G\textsubscript{2}}{G2}/M. Raw data of gene expression over time (left), with smoothed/interpolated expression over time with standard deviation (right). The expression curves for each gene have been standardized by subtracting their mean and dividing by the standard deviation over the eight time points. x-axis denotes sample time point $k$, referring to 0, 8, 16, $\dots$, 56 h after growth medium introduction. y-axis is normalized expression.}
\label{fig-cellcycle}
\end{figure*}

Gene expression for a given cell is dependent on a number of factors, including (but not limited to): cell type, cell cycle stage, circadian rhythm stage, and growth conditions. In order to best capture the natural fibroblast dynamics from population-level data, time series RNA-seq was performed on cells that were cell cycle and circadian rhythm synchronized in normal growth medium conditions (\hyperref[supp]{See SI}). Prior to data collection, all cells were temporarily held in the first stage of the cell cycle, 
\textcolor{black}{\texorpdfstring{G\textsubscript{0}}{G0}/\texorpdfstring{G\textsubscript{1}}{G1}},
via serum starvation. Upon release into the cell cycle, the population was observed every $\Delta t=8$ hours (h) for 56 h, yielding 8 time points (at 0, 8, 16, $\dots$, 56 h). Let $g_{i,k}$ be the measured activity of gene $i=1 ,\dots, N$ at measurement time $k = 1,\dots, 8$, where $N$ is the total number of human genes observed (22,083). Analysis of cell-cycle marker genes indicated that the synchronized fibroblasts took between 32-40 h to complete one cell cycle post growth medium introduction, after which cells became largely unsynchronized (Fig. \ref{fig-cellcycle}). Because of this, we define $K=5$ to be the total number of time points used for this model.

An obstacle to using $g$ to represent $x$ in a dynamical systems approach is the computational feasibility of studying a system with over 20,000 variables, necessitating a dimension reduction. Na\"{i}ve dimension reductions such as partitioning the genome into 1 mega-base pair (Mb) bins ignores inherent structural organization of the genome and obscures important intricacies of finer resolutions. A comprehensive genome state representation should include aspects of both structure and function, and simultaneously have low enough dimension to be computationally reasonable. Along these lines, we propose a biologically inspired dimension reduction based on topologically associated domains (TADs).

\begin{figure}[h]
\centering
\includegraphics[width = .49\textwidth]{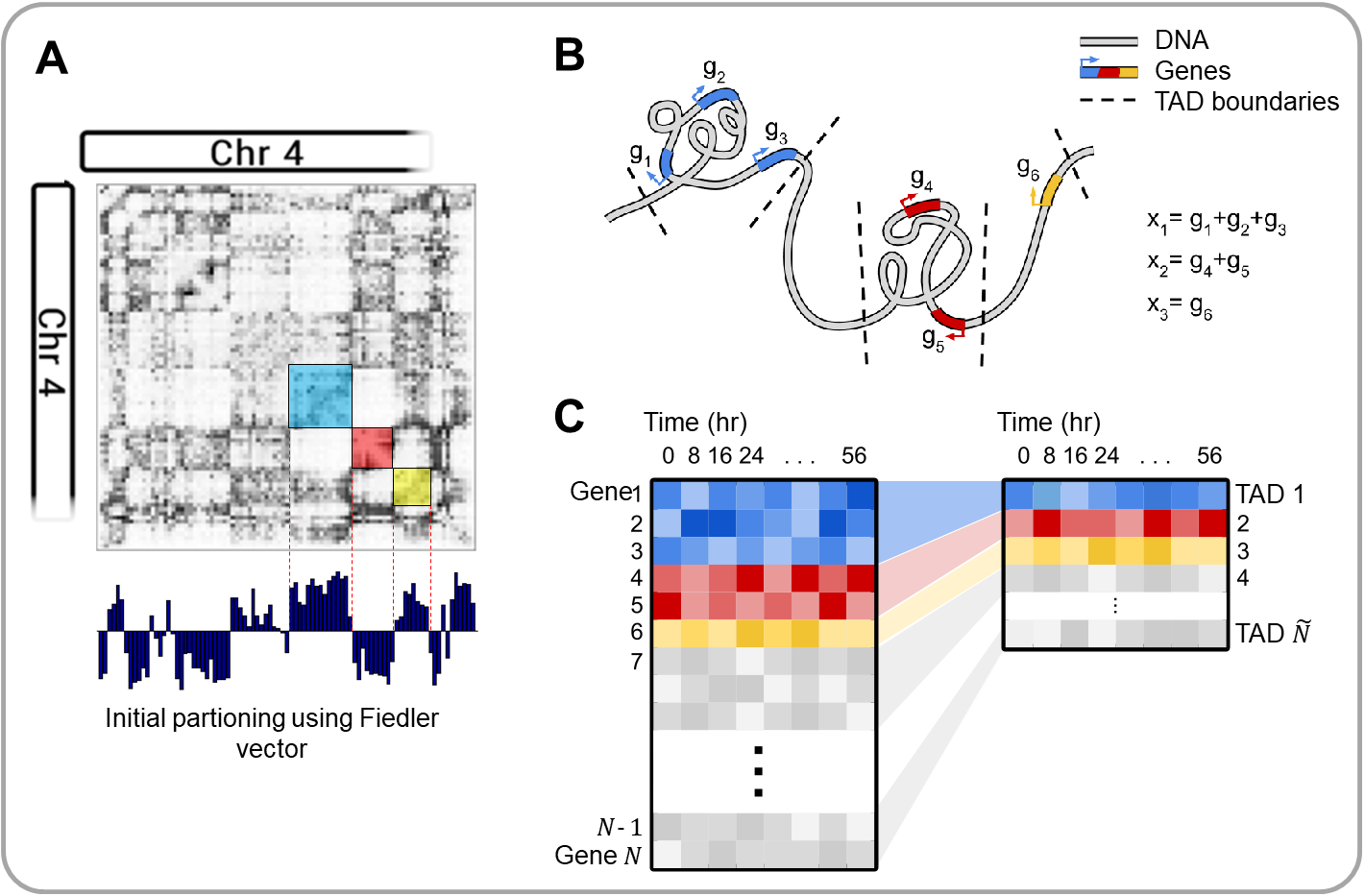}
\caption{\label{TAD_cartoon} Overview of TAD dimension reduction.
(\textit{A}) Partitioning the Hi-C matrix based on the Fiedler vector.
(\textit{B}) Cartoon depiction of TAD genomic structure.
(\textit{C}) TAD dimension reduction summary.
}
\end{figure}

The advent of genome-wide chromosome conformation capture (Hi-C) allowed for the studying of higher order chromatin structure and the subsequent discovery of TADs \cite{dixon2012topological}. TADs are inherent structural units of chromosomes: contiguous segments of the 1-D genome for which empirical physical interactions can be observed \cite{chen2015functional}. Moreover, genes within a TAD tend to exhibit similar activity, and TAD boundaries have been found to be largely cell-type invariant \cite{chen2015functional,tads-ciabrelli-2015,dixon2016chromatin}. TADs group structurally and functionally similar genes, serving as a natural dimension reduction that preserves important genomic properties. Fig. \ref{TAD_cartoon} depicts 
\textcolor{black}{an overview of this concept. }
\textcolor{black}{We computed TAD boundaries from Hi-C data via an algorithm that uses Fielder vector partitioning, described in Chen \emph{et al.} (\hyperref[supp]{See SI}) \cite{Jie}.}

Let $tad(i):=j$ if gene $i$ is contained within TAD $j$. We define each state variable $x_{j,k}$ to be the expression level of TAD $j=1 ,\dots,\tilde{N}$ at time $k$, where $\tilde{N} = 2,245$ is the total number of TADs that contain genes. Specifically,  $x_{j,k}$ is defined as the sum of the expression levels of all genes within the TAD, measured in reads per kilobase of transcript per million (RPKM), i.e.
\begin{equation}
x_{j,k} := \sum_{\mathclap{\substack{i \text{ s.t.}\\ tad(i)=j}}} g_{i,k}. \label{TAD-state}
\end{equation}
\noindent The vector of all TAD activities at measurement $k$ is denoted with a single subscript $x_k\in\mathbb{R}^{\tilde{N} \times 1}, k=1,\dots, K$. 

\subsection*{State Transition Matrix: $\mathbf A_k$}
Given the data we have, perhaps the most direct way to model the evolution of TAD activity level would be to assume a model of the form $x_{k+1}=x_k + y$, where $x_k$ and $x_{k+1}$ come from data, and $y$ is the vector difference of $x_{k+1}$ and $x_k$. However, the data could also be viewed in a different way. Taken over a full cycle, the  average value of the expression level of the 2,245 TADs is known, within experimental error. Assuming that there is a function $f$ which maps $x_k$ to $x_{k+1}$, we can subtract the steady state average, $\bar{x}$, and focus on measuring the deviation from average as the cycle evolves. With this in mind, we have $f(x) =\bar{x} + A(x-\bar{x})$ where $A$ is allowed to depend on $x$'s location in the cell cycle. That is, we build a model for the variation of the cell cycle about $\bar{x}$. For the model to match data and capture variability over the cell cycle, we will need to have a different $A$ for each time step. Using the principle that $A$ should differ as little from the identity as  possible, we let $A_k$ be the identity plus a rank one matrix chosen to match the data, for each time step $k$. In this case we have $x_{k+1} -\bar{x} = A_k(x_k-\bar{x})$.

We define a time dependent state transition matrix $A_k$.
\begin{align}
A_k &:= I_{\tilde{N}} + \frac{(x_{k+1}-x_k)x_k^\transposeSR}{x_k^\transposeSR x_k} \in\mathbb{R}^{\tilde{N}\times \tilde{N}}, ~ ~ ~ ~ k=1,2,3,4,5
\end{align}
\noindent where $I_{\tilde{N}}$ is the $\tilde{N}\times \tilde{N}$ identity matrix.

\subsection*{Input Matrix and Input Signal: $\mathbf B, \mathbf u_k$}

With the natural TAD-level dynamics established in the context of our control Eq. \ref{eqn-linear}, we turn our attention to quantifying methods for control.

A TF is a protein that can regulate a gene positively or negatively by binding to a specific DNA sequence near a gene and encouraging or discouraging transcription. This is accomplished, for example, by altering local chromatin conformation or by recruiting RNA polymerase II and other transcriptional machinery \cite{Latchman-1997}. The degree to which a TF activates or represses a gene depends on the specific TF-gene interaction and most likely on a variety of nuclear subtleties and intricacies that are difficult to quantify. Let $w_{i,m}$ be the theoretical \emph{regulation weight} of TF $m$ on gene $i$, where $w_{i,m}>0$ ($w_{i,m}<0$) if TF $m$ activates (represses) gene $i$, and $m = 1,\dots, M$, where $M$ is the total number of well-characterized TFs. Weights that are bigger in absolute value, $|w_{i,m}|\gg 0$, indicate stronger transcriptional influence, and weights equal to zero, $w_{i,m}=0$, indicate no influence.

\begin{figure*}
\centering
\includegraphics[width=.85\textwidth]{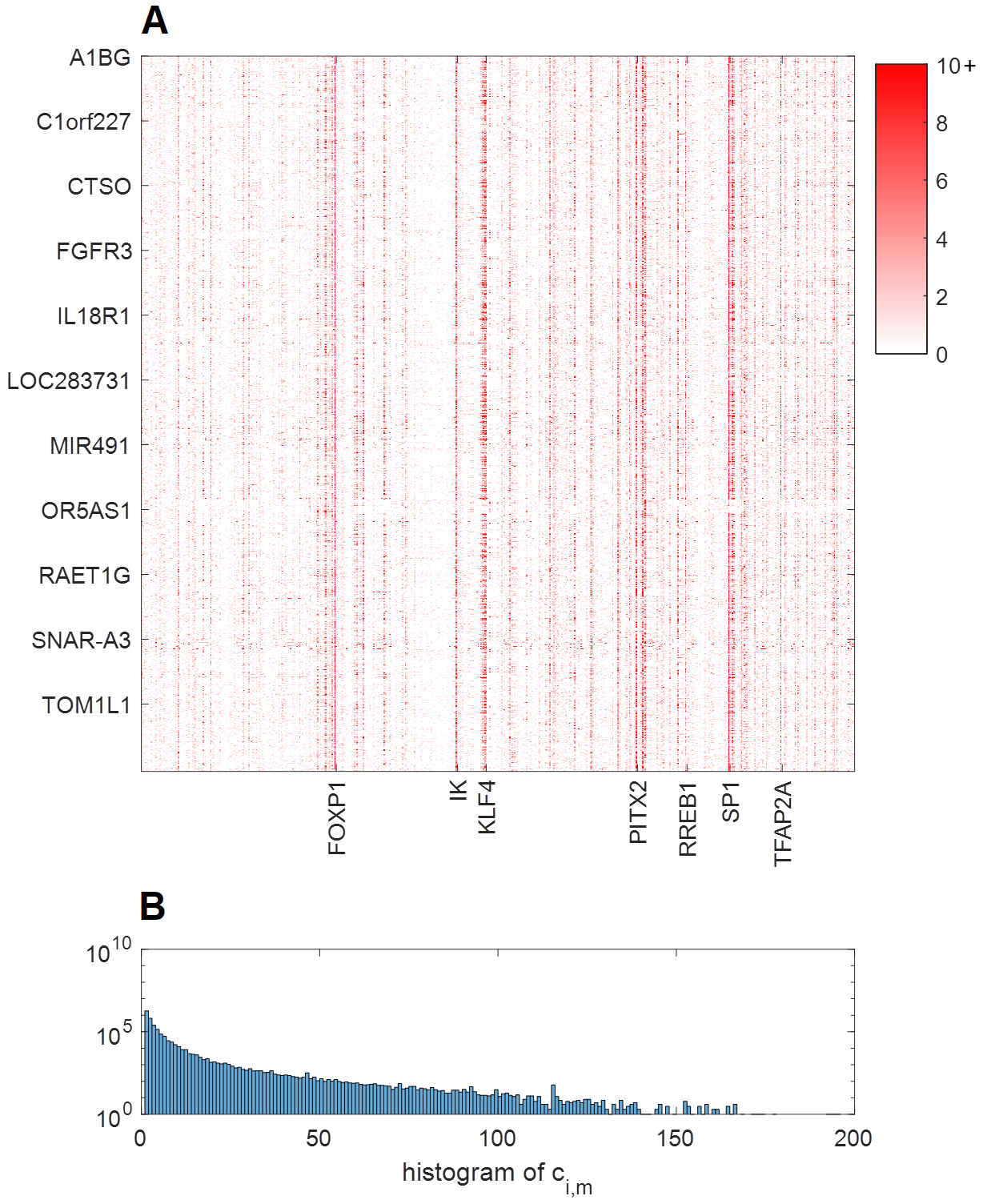}
\caption{Visualization of input matrix $\mathbf B$.
(\textit{A}) Visualization of the 22,083$\times$547 $c_{i,m}$ matrix: identified TF-to-gene interactions based on TFBSs. The color at entry $(i,m)$ represents how many transcription factor $m$ TFBSs were observed within $\pm5$kb of gene $i$'s TSS. The color axis has been truncated to $[0,10]$ but note that more than 10 TFBSs were observed for many (gene,TF) pairs. Certain columns (TFs) are dramatically highlighted compared to others, some of which have been labeled by name along the horizontal axis. Some gene names are labeled along the vertical axis, none of which particularly stand out. Both genes and TFs are sorted alphabetically.
(\textit{B}) A histogram for the non-zero values of $c_{i,m}$. The log-scale on the vertical axis emphasizes that most of the gene TSS regions contain much less than 25 TFBSs for a given TF. The \textit{SP1} TFBS, for example, is observed 249 times in a 10kb TSS centered on a gene.}
\label{fig-consensus}
\end{figure*}

Extensive TF perturbation experiments would be needed to determine $w_{i,m}$ for each TF $m$ on each gene $i$. Instead, we propose an alternative (simplified) method to approximate $w_{i,m}$ from existing, publicly available data for TF binding sites, gene accessibility, and average activator/repressor activity. 
\textcolor{black}{To determine the number of possible binding sites a TF $m$ recognizes near gene $i$, we scanned the reference genome for the locations of potential TF binding sites (TFBSs) (\hyperref[supp]{See SI}).} 
Position frequency matrices (PFMs), which give information on TF-DNA binding probability, were obtained for 547 TFs from a number of publicly available sources ($\therefore M = 547$). Let $c_{i,m}$ be the number of TF $m$ TFBSs found within $\pm5$kb of the transcriptional start site (TSS) of gene $i$ (Fig. \ref{fig-consensus}). In our model, the magnitude of $w_{i,m}$ is proportional to $c_{i,m}$. False negatives would include distal TFBSs outside of the $\pm5$kb window, while false positives would be erroneous TFBS matches.

Although many TFs can do both in the right circumstances, most TFs have tendency toward either activator or repressor activity \cite{Latchman-1997}. That is, if TF $m$ is known to activate (repress) most genes, we can say with some confidence that TF $m$ is an activator (repressor), so $w_{i,m}\geq 0$ ($w_{i,m}\leq 0$) for all $i$. To determine a TF's function, we performed a literature search for all 547 TFs and labeled 299 as activators and 124 as repressors (\hyperref[supp]{See SI}). The remaining TFs were labeled unknown for lack of conclusive evidence for activator or repressor function. In the case of inconclusive evidence, the TF was evaluated as both an activator and a repressor in separate calculations. Here, we define $a_m$ as the activity of TF $m$, with 1 and -1 denoting activator and repressor, respectively, and the sign of $w_{i,m}$ will be determined by $a_m$.

\textcolor{black}{TFBSs} are cell-type invariant since they are based strictly on the linear genome. However, it is known that for a given cell type, certain areas of the genome may be opened or closed depending on epigenetic aspects. To capture cell type specific regulatory information, we obtained gene accessibility data through DNase-seq. DNase-seq extracts cell type specific chromatin accessibility information genome-wide by testing the genome's sensitivity to the endonuclease DNase I, and sequencing the non-digested genome fragments. This data is used for our initial cell type to determine which genes are available to be controlled by TFs \textcolor{black}{\cite{thurman2012accessible}}. Here, we define $s_i$ be the DNase I sensitivity information (accessibility; open/close) of gene $i$ in the initial state, with 1 and 0 denoting accessible and inaccessible, respectively (\hyperref[supp]{See SI}).

We approximate $w_{i,m}$ as
\begin{equation}\label{eq: w_def}
w_{i,m} := a_m s_i c_{i,m},
\end{equation}
\noindent so that the magnitude of influence is equal to the number of observed consensus motifs $c_{i,m}$, except when the gene is inaccessible ($s_i$ = 0) in which case $w_{i,m}=0$.

Since we are working off a TAD-dimensional model, our input matrix $B$ must match this dimension. Let $b_m$ be a 2,245-dimensional vector, where 
the $j^{th}$ component is 
\begin{equation}\label{eq: b_gene2TAD}
b_{j,m} := \sum_{\mathclap{\substack{i \text{ s.t.}\\ tad(i)=j}}} w_{i,m} 
\end{equation}
and define a matrix $B= \left[ \begin{array}{cccc} b_1\; b_2\; \cdots b_M\end{array}\right]$.

The amount of control input is captured in $u_{k}$, which is a $\mathbb{R}^{M \times 1}$ vector representing the quantity of the external TFs we are inputting to the system (cell) at time $k$. This can be controlled by the researcher experimentally through manipulation of the \textcolor{black}{TF concentration} \cite{Brewster-2014}. In this light, we restrict our analysis to $u_{k}\geq 0$ for all $k$, as TFs cannot be subtracted from the cell. $u_{m,k}$ is defined as the amount of TF $m$ to be added at time point $k$. 

\begin{figure*}[h]
\begin{center}
\includegraphics[width=0.9\textwidth]{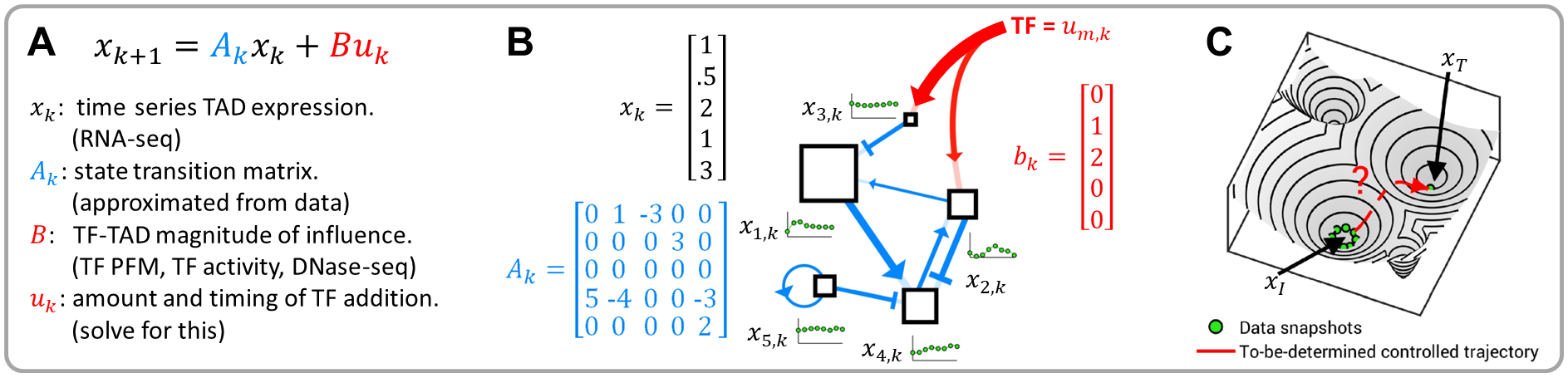}
\caption{\label{fig-tad-network} Data-guided control overview.
(\textit{A}) Summary of control equation variables.
(\textit{B}) Each TAD is a node in a dynamic network. The blue connections represent the edges of the network and are determined from time series fibroblast RNA-seq data. The miniature green plots represent the expression of each TAD changing over time. The red arrows indicate additional regulation imposed by exogenous transcription factors.
(\textit{C}) A conceptual illustration of the problem: can we determine transcription factors to push the cell state from one basin to another?}
\end{center}
\end{figure*}

With all variables of our control Eq. \ref{eqn-linear} defined, we can now attempt to predict which TFs will most efficiently achieve cellular reprogramming from some $\xinitial$ (initial state; fibroblast in our setting) to $\xtarget$ (target state; any human cell type for which compatible RNA-seq data is available) through manipulation of $u_{k}$. An overview of our DGC framework is given in Fig. \ref{fig-tad-network}.

\subsection*{Selection of TFs}

We consider different scenarios for the type of input regime. The first is assuming the input signal is constant $u_1=u_k=\bar{u}$, intended to mimic empirical regimes where TFs are given at a single time point. We will show that this method is theoretically inferior to inputting different TFs at different points in the cell cycle in a later section.

Eq. \ref{eqn-linear} has an explicit solution that is easily computed.
\begin{align*}
x_2 &= A_1x_1 + Bu_1 \\
x_3 &= A_2A_1x_1 + A_2Bu_1 + Bu_2 \\
x_4 &= A_3A_2A_1x_1 + A_3A_2Bu_1 + A_3Bu_2 + Bu_3 \\
 &\vdots 
\end{align*}
\noindent Notice the expression for $x_4$ depends the input matrix $B$ and the input signal $u_k$. 

If $\xtarget$ is a target condition, then the Euclidean distance $\|\cdot\|$ can be used to measure how close a state is to the target state, i.e.:
\begin{equation}
d = \| \xtarget - x_6(u_k) \|,
\end{equation}
\noindent where the notation $x_6(u_k)$ is used to emphasize the dependence of $x_6$ on $u_k$. Considering all possible input signals, one can compute the optimal control that finds the minimum distance for a given initial and target cell type. 
\textcolor{black}{Let $u_{*k}$ denote the optimal $u_{k}$ used to minimize $d$, and $d_*$ denotes this minimum distance value.}

We note here that when determining which TFs can be used to reach $\xtarget$, it is often desirable and more experimentally feasible to minimize the number of distinct TFs given to the cells. Transfection of cells with multiple different TFs can lead to cell stress and death, and a lower efficiency of transfection overall. Moreover, many experimentally confirmed direct reprogramming regimes use $\leq$4 TFs to achieve reprogramming. For these reasons, we set all indices of $u_k$ equal to zero, except for indices corresponding to TFs that we choose.

We define $\hat p$ to be a set of positive integers that refer to the indices of $u_k$ (read: TFs) that are allowed to be non-zero (e.g. $\hat p = \{1,4,7\}$ refers to TFs 1, 4, and 7). Let $p$ be the number of elements in $\hat p$. Given a set of TFs ($\hat p$), we determine the quantity and timing of TF input ($u_{*k}$) 
that minimizes 
the difference ($d_*$) between the initial ($\xinitial$) and target ($\xtarget$) cell state. Mathematically, this can be written as
\color{black}
\begin{align}
\begin{array}{ll}
\displaystyle \minimize_{u_k} & \| \xtarget-x_6(u_k) \| \\
\stj & \begin{cases}
    u_{m,k} \geq 0,& \text{if } m\in \hat p\\
    u_{m,k} = 0,& \text{if } m\not\in \hat p
\end{cases}
\end{array}
\label{eq: prob_u_}
\end{align}
\color{black}
\noindent We use MATLAB's \textit{lsqnonneg} function to solve Eq. \ref{eq: prob_u_}, which gives $u_{*k}$ and $d_*$.

Let $d_0 := \|\xtarget - \xinitial \|$, be the distance between initial and target states with no external influence. We define the score $\mu := d_0 - d_*$, which can be interpreted as the \emph{distance progressed towards target}. $\mu$ can be calculated for each $\hat p$ and sorted (high to low) to determine which TF or TF combination is the best candidate for direct reprogramming between $\xinitial$ and $\xtarget$.

\vspace{10pt}
\noindent \textbf{Remark:} Subsets of TFs were chosen for each calculation based on the following criteria: $\geq$10-fold expression increase in target state compared to initial state, and $\geq$4 RPKM in target state. These criteria are used to select differentially expressed TFs and TFs that are sufficiently active in the target state.

\section*{Results}
\subsection*{Quantitative Measure Between Cell Types}
In order to best utilize our algorithm to predict TFs for reprogramming, compatible data on target cell types must be collected. For this, we explore a number of publicly available databases where RNA-seq has been collected, along with RNA-seq data collected in our lab. The ENCODE Consortium has provided data on myotubes and embryonic stem cells (ESCs) (\hyperref[supp]{See SI}) \cite{encode2012integrated}. The GTEx portal provides RNA-seq data on a large variety of different human tissue types \cite{lonsdale2013genotype}. Although each GTEx experiment is performed on tissue samples, thus containing multiple different cell types, we use these data as more general cell state targets.

\begin{figure*}[h]
\includegraphics[width = \textwidth]{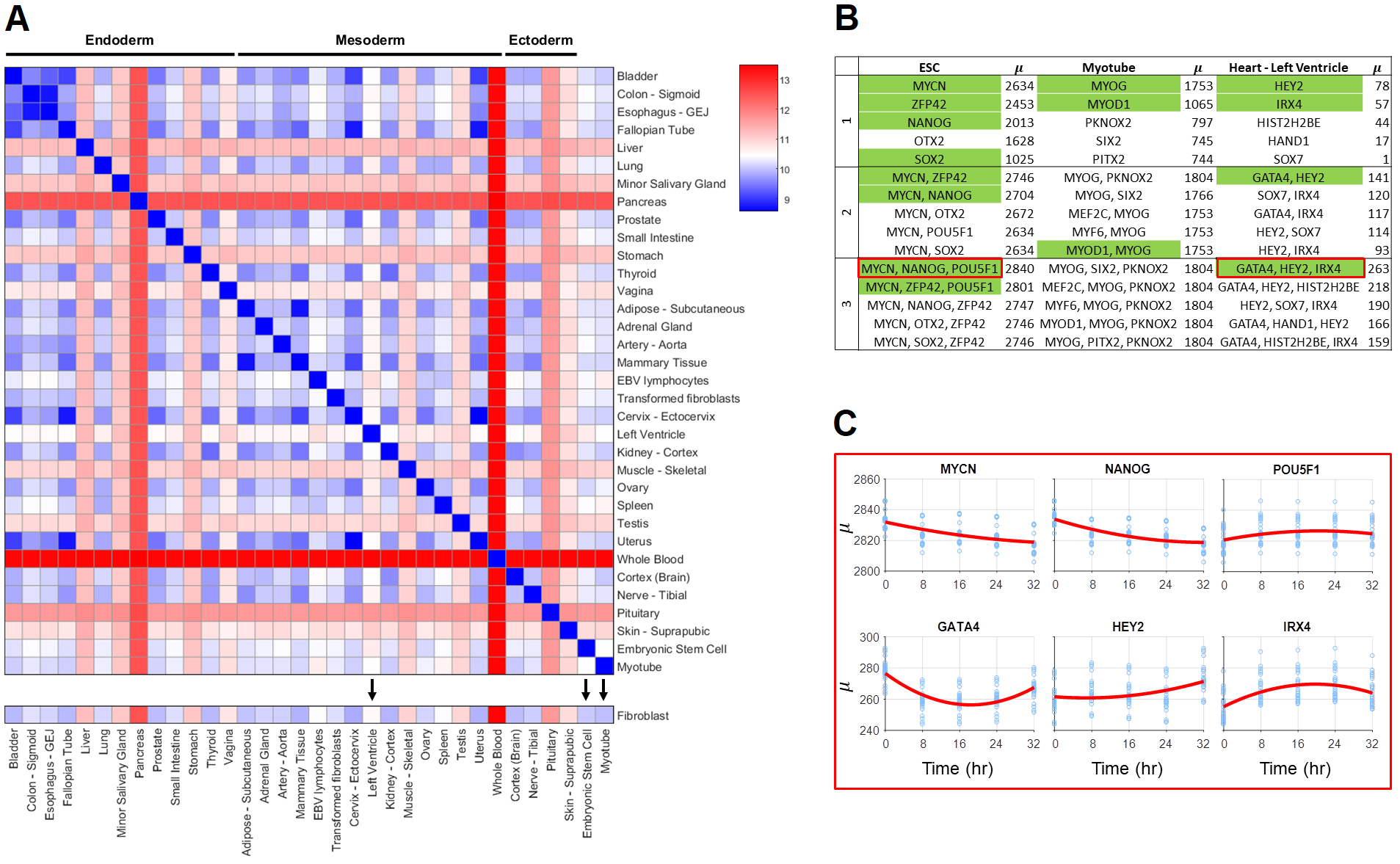}
\caption{Quantitative measure between cell types and transcription factor scores.
(\textit{A}) $d_0$ values between GTEx tissue types and ESC, myotube, and fibroblast. Tissue types and cell types with black arrows have predicted transcription factors for reprogramming from fibroblasts shown in \ref{dist_mat}B.
(\textit{B}) Table of predicted transcription factors for a subset of cell and tissue types. Top 5 transcription factors for combinations of 1-3 shown. Green labeled transcription factors are either highly associated with the differentiation process of the target cell type and/or validated for reprogramming. These transcription factors are discussed in the main text.
(\textit{C}) Time-dependent scores for selected combinations of 3 transcription factors for fibroblast to ESC and fibroblast to ``Heart - Left ventricle". x-axis refers to time of transcription factor addition, y-axis refers to $\mu$.
} \label{dist_mat}
\end{figure*}

To give a numerical structure to cell type differences, conceptually similar to Waddington's epigenetic landscape, we calculate $d_0$ between all cell types collected. Fig. \ref{dist_mat}\textit{A} shows $d_0$ values for 32 tissue samples collected from the GTEx portal, along with ESC, myotube, and our fibroblast data (additional cell type $d_0$ values shown in \hyperref[supp]{SI}). Warmer colors (red) denote further distances between cell types. GTEx RNA-seq data is 
scaled 
to keep total RPKM difference between time series fibroblast and GTEx fibroblast RNA-seq minimal (\hyperref[supp]{See SI}).

\subsection*{TF Scores}
To assess our method's predictive power, a subset of target cell types are presented here that have either validated TF reprogramming methods or TFs highly associated with the target cell type. Additional predicted TFs for reprogramming are included in \hyperref[supp]{SI}. 
We note that though experimentally validated TFs provide the best current standard for comparison, we believe experimental validation with our predicted TFs may provide more efficient reprogramming results. For all reprogramming regimes presented in this section, fibroblast is used as the initial cell type due to the availability of synchronized time series data, and all TFs are introduced at $k=1$ \cite{chen2015functional}.

For conversion of fibroblast to myotubes, the top predicted single input TFs are MYOG and MYOD1, both of which are known to be crucial for myogenesis. While MYOD1 is the classic master regulator reprogramming TF for myotube conversion, activation of downstream factor MYOG is necessary for full conversion \cite{weintraub1993myod}. For fibroblast to ESC conversion, a number of TFs known to be necessary for pluripotency are predicted, including MYCN, ZFP42, NANOG, and SOX2 \cite{takahashi2007induction}. With the knowledge that no single TF has been shown to fully reprogram a fibroblast to an embryonic state, combinations of TFs are more informative for this analysis. The top scoring combination of 3 TFs is MYCN, NANOG, and POU5F1--three well-known markers for pluripotency \cite{takahashi2007induction}. Interestingly, POU5F1 scores poorly when input individually, but is within the top set of 3 TFs when used in combination with MYCN and NANOG. Left ventricle reprogramming includes TFs that are known to be necessary for natural differentiation in the top score for all 1-3 combinations. These include GATA4 (a known TF in fibroblast to cardiomyocyte reprogramming), HEY2, and IRX4 \cite{ieda2010direct,fischer2004notch,nelson2014irx4}.

\subsection*{Time-dependent TF Addition}
Fibroblast to ESC conversion was of particular interest in our analysis as this is a well-studied regime with a number of validated TFs (with a variety of reported efficiencies), and this conversion is promising for its regenerative medicine application. High scoring TFs yield many that are known markers for pluripotency, but the top combination of 3, MYCN, NANOG, and POU5F1, has not been used specifically together, to our knowledge.

Since our method incorporates dynamical RNA-seq data, analysis can be extended to determine the best time to input control for a given set of TFs. In our model, there are five possible input times: 0, 8, $\dots$, 32 h. We assume a TF continues to influence the system at a constant value once input until the final time (40 h). We restrict our analysis here to combinations of 3 TFs. This gives $5^3=125$ possible \emph{Time-dependent regimes} to input the TFs; e.g. \emph{TF1, TF2, TF3} are input, respectively, at times 0,0,0, or 0,0,8, or 0,0,16, or $\dots$ or 32,32,24 or 32,32,32. Inputting a TF at time $k^*$ can be viewed mathematically as requiring $u_{m,k}=0$ for all $k<k^*$.

Time-dependent analysis of the top scoring ESC TFs reveals that scores vary widely depending on the time of input. MYCN and NANOG show a strong preference for input at the beginning of the cell cycle, while POU5F1 shows a slight preference for input towards the end of the cell cycle, with the highest score achieved when MYCN and NANOG are input at 0 h and POU5F1 is input at 32 h. Analysis on how the time of input control affects $\mu$ is shown in Fig. \ref{dist_mat}\textit{C}. Time-dependent analysis was also conducted for the top combination of 3 TFs for fibroblast to left ventricle. This set includes GATA4, HEY2, and IRX4, all factors highly associated with the cardiac phenotype \cite{ieda2010direct,fischer2004notch,nelson2014irx4}. This analysis predicted that the best reprogramming results would occur if GATA4 is given immediately (0 h), with IRX4 and HEY2 given later (24 and 32 h, respectively).

\section*{Discussion}
The results from this algorithm show promise in their prediction of known reprogramming TFs, and demonstrate the importance of including time series data for gene network dynamics. 
Time of input control has shown to have an impact on the end cell state, in line with what has been shown in natural differentiation \cite{loh2016mapping}. 

While we believe that this is the best model currently available for predicting TFs for reprogramming, we are aware of its limitations and assumptions. 
TAD-based dimension reduction is based on the observation that genes within them correlate in expression over time, though we lack definitive proof of regulation by shared transcriptional machinery \cite{chen2015functional}. This assumption was deemed necessary for dimension reduction in the context of deriving transition matrix $A_k$. With finer time steps in RNA-seq data, the assumption may not be necessary for TF prediction, at the cost of increased computation time. Additionally, a 5kb window flanking the TSS of each gene was used to ensure that all potential regulators are found, at the cost of potential inclusion of false positive motifs.

GTEx data proved to be an invaluable resource for testing our algorithm, providing many target states for TF prediction. It is important to note that these data are collected from cadaver tissue samples; therefore the RNA-seq data is coming from a heterogeneous cell-type population and may be enriched for stress factors known to be elevated after death (e.g. HSF4). Ideally, RNA-seq data for target cells would be derived from a homogeneous population, with minimal experiment collection variables. Future work includes the extension of this DGC approach to other target cell types.

Although this program can score TFs relative to other TFs in a given reprogramming regime, it is difficult to predict a $\mu$ threshold that would guarantee conversion. Additionally, rigorous experimental testing will be required to validate these findings and determine how our $u$ vector translates to TF concentration. This is a product of the large number of assumptions that must be made to develop the initial framework for a reprogramming algorithm. With finer resolution in the time series gene expression, more subtle aspects of the genomic network may be observed, allowing for better prediction.

Our proposed data-guided control framework successfully identified known TFs for fibroblast to ESC and fibroblast to muscle cell reprogramming regimes. The framework rates individual TFs as well as sets of TFs. We employ a biologically-inspired dimension reduction via TADs, a natural partitioning of the genome. This comprehensive state representation was the foundation of our framework, and the success of our methods motivates further investigation of the importance of TADs as functional units to control the genome.

A dynamical systems view of the genome allows for analysis of timing, efficiency, and optimality in the context of reprogramming. Our framework is the first step toward this view. The successful implementation of time-varying reprogramming regimes would open new avenues for direct reprogramming. Experimental verification of predicted regimes and development of methods to identify optimal sets of TFs are planned for the near future. This template can be used to develop regimes for changing any cell into any other cell, for applications that include reprogramming cancer cells and controlling the immune system. Our DGC framework is well equipped for designing personalized cellular reprogramming regimes. Finally, this framework can serve as a general technique for investigating the controllability of networks strictly from data.

\section*{Methods and Materials}
Hi-C and RNA-seq data were collected from cell cycle and circadian rhythm-synchronized proliferating human fibroblasts of normal karyotype. Data were collected every 8 h, spanning 56 h. Publicly available data was used for target cell types. Detailed materials and methods are provided in Chen \emph{et al.} and in \hyperref[supp]{SI} \cite{chen2015functional}.

\section*{Acknowledgement}
We thank Robert Oakes, Emily Crossette, and Sijia Liu for their critical reading of the manuscript and helpful discussions. We extend special thanks to James Gimlett and Srikanta Kumar at Defense Advanced Research Projects Agency (DARPA) for support and encouragement. This work is supported, in part, by the DARPA Biochronicity Program and the DARPA Deep-Purple and FunCC Program.

\bibliographystyle{unsrt}
\bibliography{data_guided_control_bib}

\newpage
\beginsupplement
\section*{Supporting Information (SI)}\label{supp}

\subsection*{Data} \label{app-dataparse}
The fibroblast (FIB) data (Hi-C and RNA-seq) used for this application was originally collected and published in a paper by Chen \emph{et al.} \cite{chen2015functional}. We refer the reader to this paper for a full description of technical protocols. 
Embryonic Stem Cell (ESC) and myotube (MT) data was downloaded from NCBI-GEO  (GSE23316 ENCODE Caltech RNA-seq and GSE52529) \cite{encode2012integrated}. 53 different tissue RNA-seq samples were downloaded from GTEx portal \cite{lonsdale2013genotype}. 51 different immune cell type RNA-seq samples were obtained from the BLUEPRINT Epigenome project \cite{adams2012blueprint}. 

\vspace{10pt}
\subsection*{Hi-C and Construction of TADs}
We computed TAD boundaries from genome-wide chromosome conformation capture (Hi-C) data using an algorithm described in Chen \emph{et al.} \cite{Jie}. The algorithm was applied to averaged time series Hi-C data from proliferating human fibroblast (FIB) at 100 kilo-base pair (kb) resolution, which identified 2,562 TADs across all autosomal chromosomes (i.e. excluding Chromsomes X and Y). Of the 2,562 TADs, 317 contained no genes and were excluded from our analysis, leaving 2,245 TADs. These TADs ranged in size from a few hundred kb to several Mb, and contained on average 9-10 genes (standard deviation of 18 genes); one gene at minimum, and 249 genes maximum. 

\vspace{10pt}
\subsection*{Construction of $\mathbf B$ Matrix}
TF binding site position frequency matrix (PFM) information was obtained from Neph \emph{et al.} and MotifDB, which is a collection of publicly available PFM databases such as, JASPAR, Jolma \emph{et al.}\, cispb\_1.02, stamlab, hPDI, UniPROBE \cite{Neph-2012,shannon2014motifdb}. TRANSFAC PFM information was included as well. Motif scanning of the human reference genome (hg19) was performed using FIMO of the MEME suite, in line with methods established by Neph \emph{et al.} \cite{Neph-2012}. DNase-seq information for human fibroblasts was derived from ENCODE for fibroblast (GSM1014531). If a narrow peak is found within the $\pm$5kb of a gene TSS, the region is classified as open. TF function information was determined through an extensive literature search. 

\vspace{10pt}
\subsection*{Scaling of RNA-seq}
Due to differences in data collection procedures, the RNA-seq RPKM values obtained from the GTEx portal were of lower value, on average, compared to our fibroblast dataset, thus favoring repressor TFs for $\mu$ scoring. In order to account for this in our model, we scaled all GTEx RNA-seq data by a factor that solves the equation

\begin{align}
\begin{array}{ll}
\displaystyle \minimize_{\mathbf \alpha} & \| \mathbf g_{FIB,UM} - \mathbf \alpha \mathbf g_{FIB,GTEX} \|
\end{array}
\end{align}

\noindent where $g_{FIB,UM}$ is the gene-level RNA-seq vector average of our fibroblast data, $g_{FIB,GTEX}$ is the gene-level RNA-seq vector of ``Cells - Transformed fibroblasts'' from the GTEx portal, and $\alpha$ is a scalar that solves this equation. For this data, $\alpha = 2.6113$ and all GTEx data used as a target state was scaled by this factor.

\vspace{10pt}
\subsection*{Removal of MicroRNA}
MicroRNA 
were removed from this analysis due to there high variance in RPKM values and unpredictable function.

\subsection*{TF Scores - Additional GTEx Data}
For fibroblast to Adipose-Subcutaneous, the highest scoring factor is EBF1, a known maintainer of brown adipocyte identity, and a known promoter of adipogenesis in fibroblasts \cite{jimenez2007critical}. The 2nd highest scoring marker, PPARG, has been shown to be involved in adipose differentiation, and can be used individually to achieve reprogramming from fibroblasts \cite{gregoire1998understanding}. Curiously, ATF3 is implicated here as being useful for adipocyte differentiation although its function has been shown to repress PPARG and stymie cell proliferation \cite{jang2012atf3}. Upon further research using time dependent addition, ATF3 addition scores best when added towards the end of reprogramming process.

2 Brain tissue samples, Cerebellum and Hippocampus, both predict TFs necessary for natural differentiation. Interestingly, our algorithm select different TFs for each conversion, with factors linked specifically to each tissue. For Cerebellum, NEUROD1, has been shown to be required for granule cell differentiation, while ZIC1 and ZIC4 are both known to promote cellebular-specific neuronal function \cite{miyata1999neurod,frank2015regulation}. The top scoring combination of 3 TFs are all similarly known to be important in neurogenesis (NEUROD1, ZBTB18, UNCX) \cite{cohen2016further,sammeta2010uncx}. Hippocampus TF scoring includes FOXG1 as the top predicted factor, a factor specifically needed in hippocampus development. OLIG2, FOXG1, and GPD1 are the top scoring set for hippocampus reprogramming, all of which have been shown to been necessary for hippocampus function.

Colon TF scoring finds known differentiation factor in natural colon secretory linage development, ATOH1, as the highest scoring individual factor \cite{vandussen2010mouse}. The top scoring combination of 2 TFs includes ATOH1 along with CDX2, another known factor necessary for full differentiation of colon cells, specifically small intestine maturation \cite{crissey2011cdx2}. Liver cell reprogramming similarly finds known factors for differentiation in the top score of all 3 combinations: HNF4A, CUX2, PROX1 \cite{delaforest2011hnf4a,seth2014prox1,vanden2005hepatomegaly}. All factor play a role in correct development of hepatic progenitor cell-types and hepatic stem cells, the cell types just above in lineage differentiation.

\subsection*{TF Scores - BLUEPRINT Project Data}

\begin{figure*}[h]
\centering
\includegraphics[width = .6\textwidth]{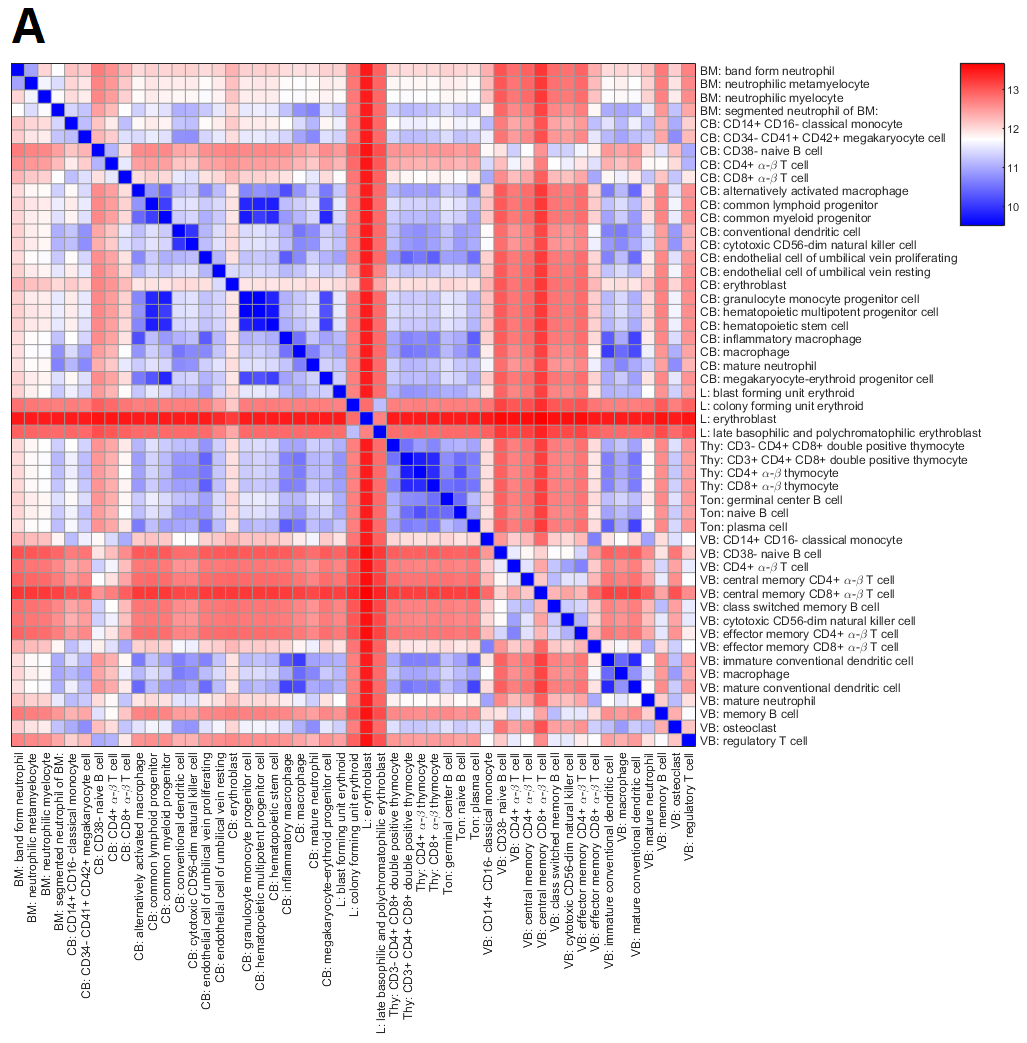}
\includegraphics[width = .9\textwidth]{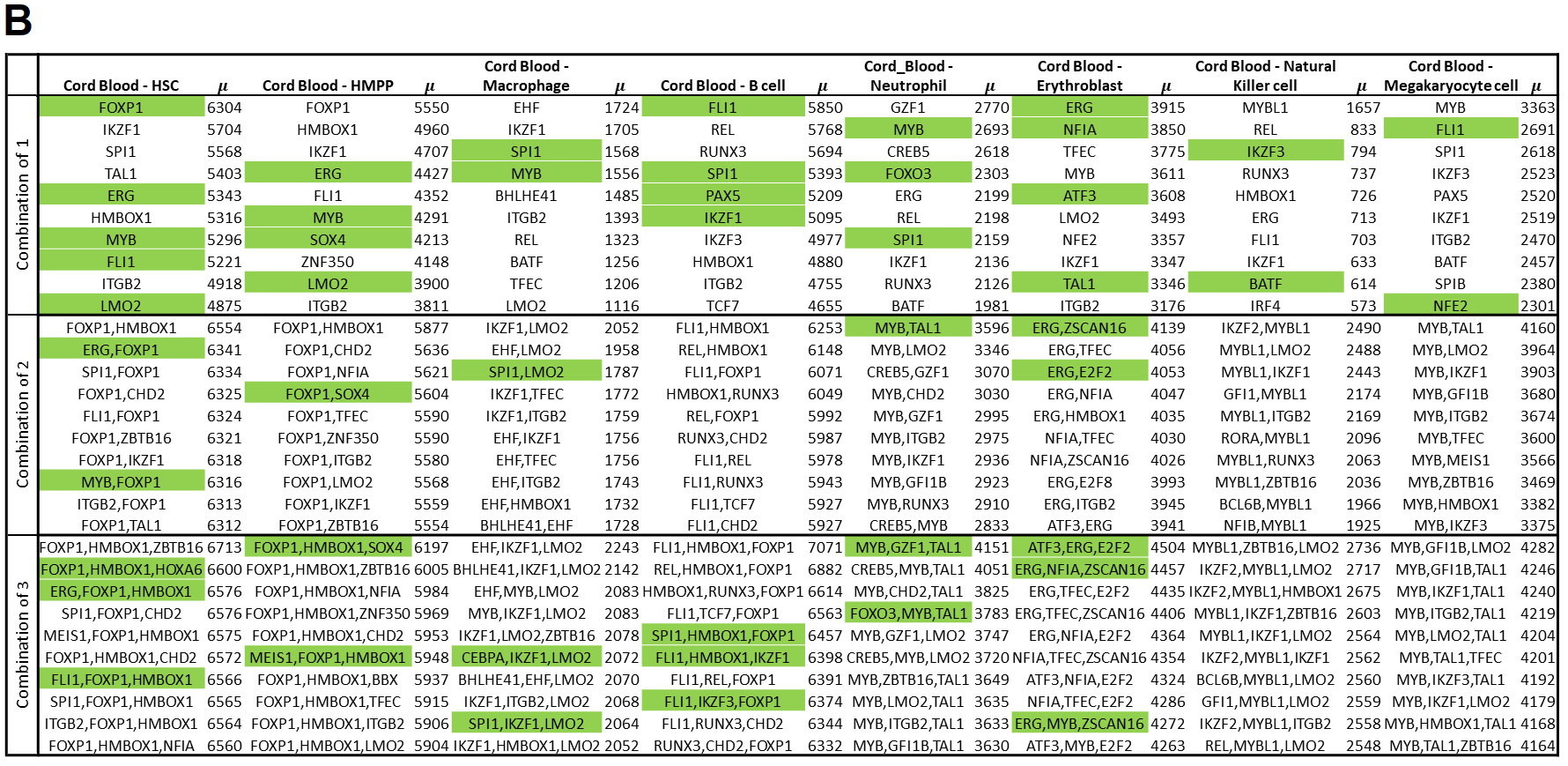}\label{TF_score_table_BP}
\caption{Quantitative measure between cell types and TF scores for BLUEPRINT Project database.
(\textit{A}) $d_0$ values between BLUEPRINT Project cell types.
(\textit{B}) TF scores for an extended list of target cell types. $\xinitial$ = fibroblast} \label{dist_mat_BP_sup}
\end{figure*}

A number of immune cell types extracted from the BLUEPRINT Project revealed promising predicted TF results when fibroblast is used as the starting point \cite{adams2012blueprint}. $d_0$ values between cell types are shown in Fig. \ref{dist_mat_BP_sup}.

For fibroblast to macrophage direct reprogramming, a number of factors scoring highly in our algorithm are known to play a role in macrophage reprogramming or the differentiation. SPI1 (along with CEBPA) has been verified experimentally to reprogram fibroblasts into macrophage-like cells \cite{feng2008pu}. IKZF1 has been shown to be crucial for macrophage polarization via the IRF4/IRF5 pathway \cite{bruns2016ikzf1}. MYB has been shown to be crucial for the upstream cell type HSC \cite{perdiguero2016development}.

For fibroblast to HSC direct reprogramming, the top scoring individual factor is highly associated with the target phenotype and has been shown to support HSC growth and regeneration \cite{naudin2017pumilio}. ERG (in combination with GATA2, LMO2, RUNX1c, and SCL) is a confirmed reprogramming factor for fibroblast to HSC in mice \cite{batta2014direct}.

For fibroblast to erythroblast reprogramming, ERG is a promising factor as it is required for the maintenance of the upstream cell type HSC \cite{batta2014direct}. NFIA is shown to promote the erythroid lineage from HSC differentiation \cite{starnes2010transcriptome}.

\subsection*{Alternative Computation of $\mathbf u$}
Below is an example of how $u$ can be computed without the constraint that $u_{k,m} \geq 0$. Assume $u_k:=\bar{u}$ is constant for all $t$. Then

\begin{equation}
x_{k+1}=A_k x_k+Bu_k. 
\end{equation}

\noindent can be written as

\begin{equation}
x_4 = A_3A_2A_1x_1 + C\bar{u},
\end{equation}

\noindent where

\[
C:= A_3A_2B + A_3B + B
\]

We seek the control $\bar{u}$ that minimizes the distance between $x(3)$ and the target $\xtarget$:

\begin{equation}
\min_{\bar{u}} \|\xtarget-A_3A_2A_1x_1 - C\bar{u} \|. \label{eqn-solveme}
\end{equation}

We can see that an exact solution exists if

\begin{equation}
\xtarget-A_3A_2A_1x_1 \in \text{span}(C), \label{eqn-spanner}
\end{equation}

\noindent and is given by 

\begin{align}
A_3A_2A_1x_1 + C\bar{u} &= \xtarget \\
C\bar{u} & = \xtarget - A_3A_2A_1x_1 \\
\bar{u} &= C^{\dagger}\left(\xtarget - A_3A_2A_1x_1\right), \label{eqn-minubar}
\end{align}

\noindent where $C^{\dagger}$ denotes the Moore-Penrose pseudoinverse of the matrix $C$, computed using the singular value decomposition of $C$. 
Even when Eqn (\ref{eqn-spanner}) is not satisfied, it is well established that the control (\ref{eqn-minubar}) solves (\ref{eqn-solveme}) . 

Define

\begin{equation}
d = \bigg\| (I_N - CC^{\dagger})(\xtarget-A_3A_2A_1x_1) \bigg\|
\end{equation}
\begin{equation}
\mu := d_0 - d_*
\end{equation}

\noindent $\mu$ can be used to compare between potential TFs for a defined initial state ($\xinitial$), target state ($\xtarget$), and TF number ($p$). The larger the value of $\mu$, the higher the relative score for its corresponding TF set.

We note that accurate TF predictions for some desired target cell type may not depend on minimizing distance alone, but also the amount of ``energy" required for the system to reach $d_*$. We denote energy here with $\mu_2$ and define it as:

\begin{equation}\label{eq: mu_2_def}
e(u) = \sum_{k=0}^{K-1} u_k\transpose\cdot u_k = \mu_2.
\end{equation}

$\mu_2$ is analogous to the amount of a TF that needs to be added to the system to achieve $d_*$. In the case where two different TFs achieve the same $\mu$ score, $\mu_2$ can be computed to decide the better candidate (i.e. lower $\mu_2$ is better TF).

\newpage

\newpage
\subsection*{SI Figures}\label{supp_figs}

\clearpage
 
\clearpage

\begin{figure*}[h]
\centering
\includegraphics[width = .6\textwidth]{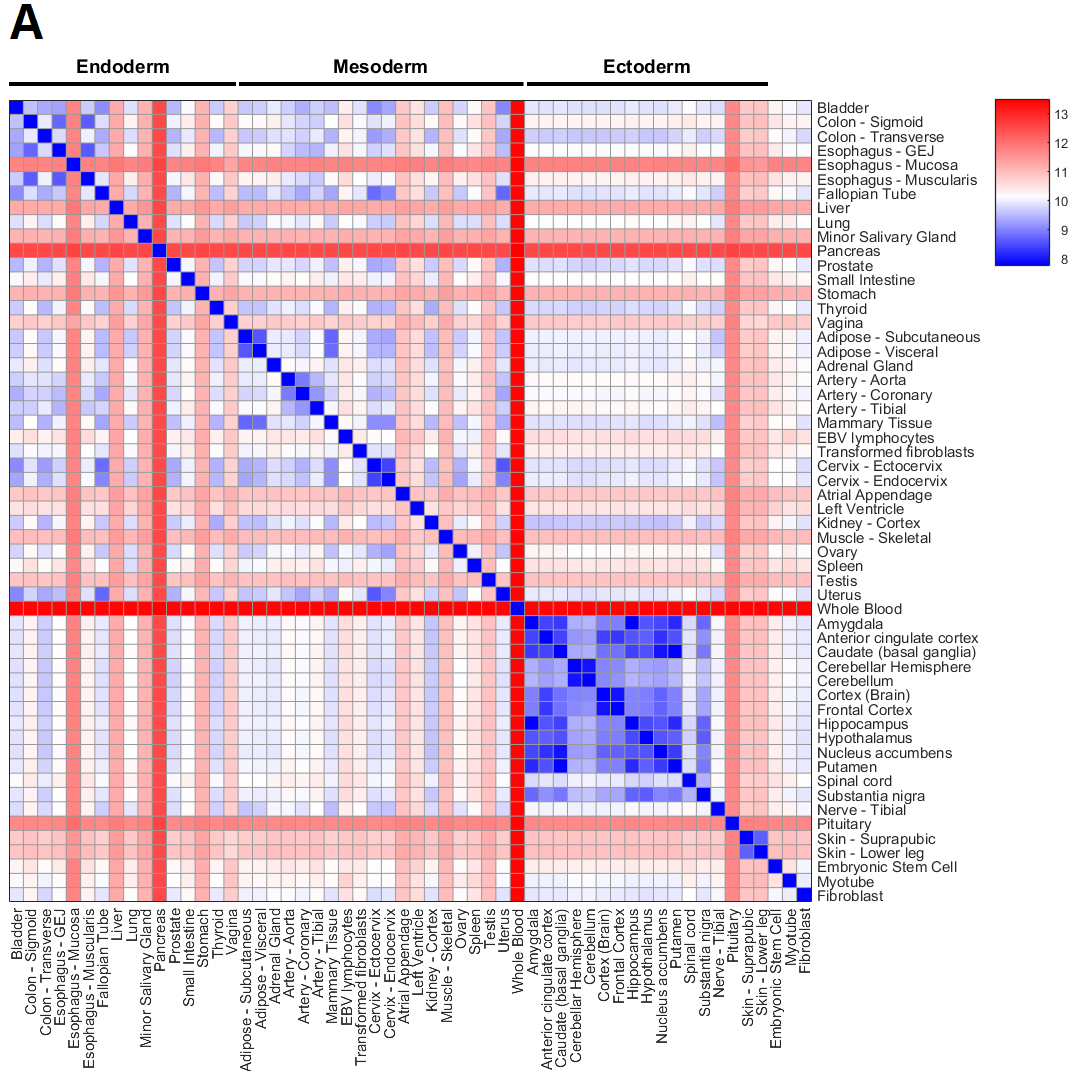}
\includegraphics[width = .9\textwidth]{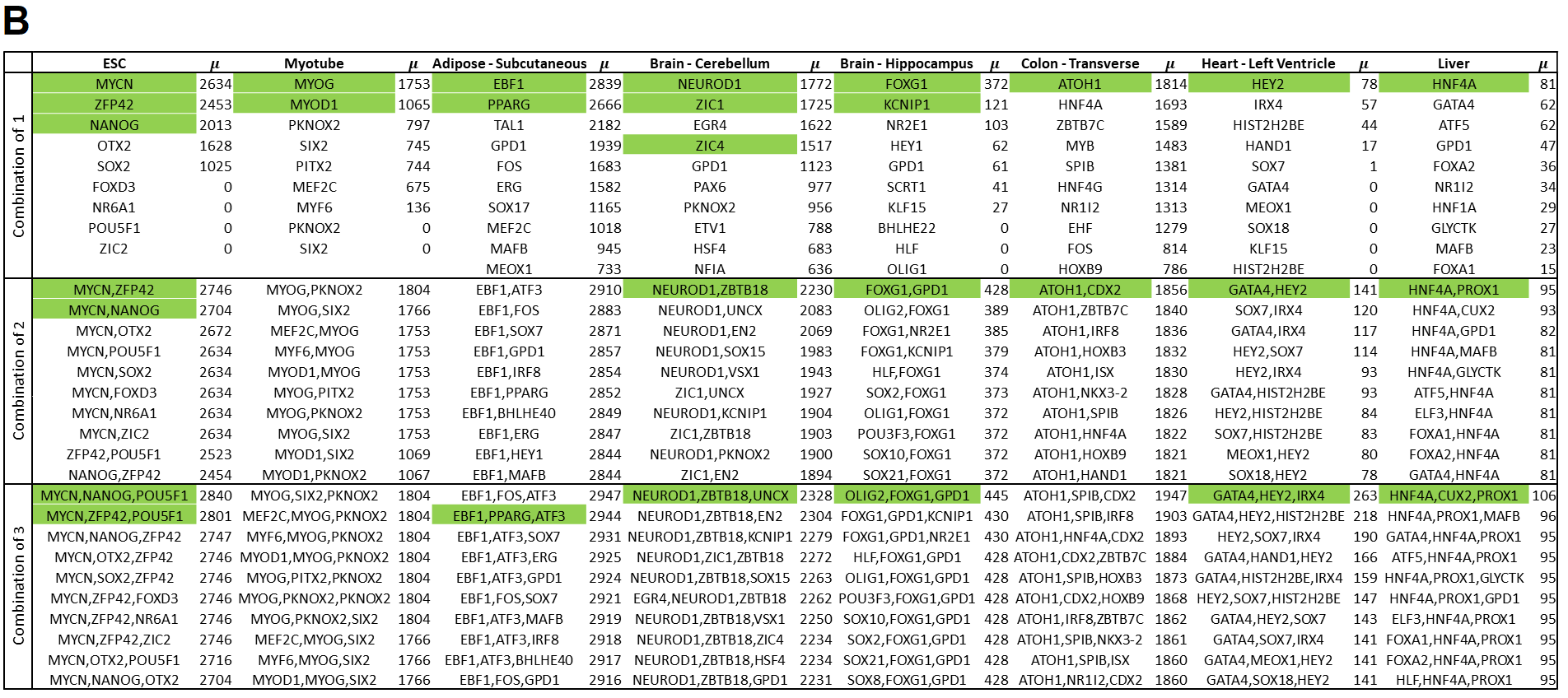}
\caption{Quantitative measure between cell types and transcription factor scores.
(\textit{A}) $d_0$ values between all GTEx tissue types.
(\textit{B}) TF scores for an extended list of target cell types. $\xinitial$ = fibroblast}
\end{figure*}

\newpage

\end{document}